\documentclass[epj]{svjour}

\usepackage{epsfig}

\usepackage{amsmath}

\begin{document}

\title{Q-Value for the Fermi Beta-Decay of $^{46}$V}

\author{T.~Faestermann\inst{1}\fnmsep\inst{2}\fnmsep\thanks{thomas.faestermann@ph.tum.de} \and 
R.~Hertenberger\inst{2}\fnmsep\inst{3}\and 
H.-F.~Wirth\inst{2}\fnmsep\inst{3}\and 
R.~Kr\"ucken\inst{1}\fnmsep\inst{2}\and 
M.~Mahgoub\inst{1}\fnmsep\inst{2}\and 
P.~Maier-Komor\inst{1}\fnmsep\inst{2}}

\institute{Physik Department E12, Technische~Universit\"at
M\"unchen,~D85748~Garching,~Germany\and
Maier-Leibnitz-Laboratorium der M\"unchner Universit\"aten (MLL),~D85748~Garching,~Germany\and
Department f\"ur Physik, Ludwig-Maximilians-Universit\"at M\"unchen,~D85748~Garching,~Germany}

\date{\today}

\abstract{
By comparing the Q-values for the $^{46}$Ti($^{3}$He,t)$^{46}$V 
and $^{47}$Ti($^{3}$He,t)$^{47}$V reactions to the isobaric analog states the Q-value for the 
superallowed Fermi-decay of $^{46}$V has been determined as
Q$_{EC}(^{46}$V$)=(7052.11\pm0.27)~keV$. The result is compatible 
with the values from two recent direct mass
measurements but is at variance with the previously most precise reaction Q-value.
As additional input quantity we have determined the neutron separation
energy S$_n(^{47}$Ti$)=(8880.51\pm0.25)~keV$.
}


\maketitle    

\section{Introduction}
For the determination of the first element $V_{ud}$ of the Cabib\-bo-Kobayashi-Maskawa (CKM) matrix,
that describes the mixing of different quark flavours,
the beta transition rates between even-A isobaric analogue states (IAS) with spin I=0,
called superallowed Fermi-transitions, yield
the most precise values up to now \cite{hardy08}.
The values from neutron or pion decay are not competitive yet \cite{PDG08}. 
For nine T=1 pairs with mother nuclei from $^{10}$C to $^{54}$Co the relevant 
quantities half-life, branching ratio and decay energy,
that enters into the phase space factor with the fifth power,
have been measured with great precision and yield with the appropriate corrections
a decay strength which is constant within the uncertainties of a few $10^{-4}$.
With the same level of precision the hypothesis of the Conserved Vector Current (CVC) in 
weak nuclear decays is proven. From the average beta-decay strength 
the value of $V_{ud}=0.97408\pm 0.00026$ has been deduced \cite{eron08} and 
with this value and the recently improved value of $V_{us}$ and of $V_{ub}$ 
the unitarity relation for the first row of the CKM matrix is now well
fulfilled with 
\[ \sum_{\nu=d,s,b}|V_{u\nu}|^2=0.9998\pm0.0010 .\]

These new measurements on $V_{us}$ in Kaon decays have now removed
the "violation" of the unitarity, which had persisted for many years (e.g. \cite{hardy05}).
Nevertheless, it is essential to determine the nuclear physics
quantities, experimental and theoretical, in independent ways,
to remove systematic errors in $V_{ud}$.
The beta-decay Q-values were traditionally determined using
nuclear reaction Q-values. 
In recent years the precision of direct mass measurements
using Penning traps has improved that much, that mass differences of radioactive
ions can be determined with the required accuracy.
Recently the decay energy of $^{46}$V, one of these Fermi-emitters, has been measured 
by direct mass 
measurements of $^{46}$V and $^{46}$Ti in the Canadian Penning Trap \cite{Sav05} 
at Argonne Nat. Lab. as well as in the JYFLTRAP \cite{eron06} at the University
of Jyv\"askyl\"a. 
Their result is at variance with the 1977 result obtained by 
Vonach et al. \cite{Von77} with a Q value measurement of the 
$^{46}$Ti($^{3}$He,t)$^{46}$V reaction. 
Savard et al. argue that all seven Q-values of Vonach et al. are erroneous and 
discard them in the averages of input data. 
Hardy et al. \cite{hardy06} have undertaken a detailed re-analysis of
(n,$\gamma$) and (p,$\gamma$) data in order to search for systematic
differences between reaction Q-values and mass differences.
Since reaction Q-value measurements seem to be regarded with scepticism, 
we repeated the $^{46}$Ti($^{3}$He,t)$^{46}$V measurement to clarify, whether there is a 
principal problem.
\begin{table}[t]
\centerline{
\begin{tabular}{|c|c|c|c|c|c|}
\hline
target&nat&46&46/47&47&48
\\\hline
isotope& & & & &  
\\\hline
46	&8.3&70.8&38.7&2.5&0.2\\
47	&7.4&3.2&37.5&76.1&0.3\\
48	&73.7&23.0&19.0&19.0&99.1\\
49	&5.4&1.6&1.5&1.3&0.2\\
50	&5.2&1.4&1.3&1.1&0.2
\\\hline
\end{tabular}
}
\caption{Isotopic composition (in atom \%) of the targets used.}
\label{tab_Q}
\end{table}

\section{Experimental Details}
We used essentially the same experimental components as Vonach et al. \cite{Von77},
the Munich Tandem accelerator and the high resolution Q3D spectrograph \cite{Q3D73}.
But instead of calibrating the ion energies with an over $100~m$ long time-of-flight
system \cite{Von77}, that is not operational any more,
we used the same reaction on another Ti isotope within the same target
as a calibration. This method may proof to be reliable also in other cases,
since most systematic uncertainties are avoided.
\begin{figure}[b,t]
\centerline{\includegraphics[width=9.5cm,angle=0]{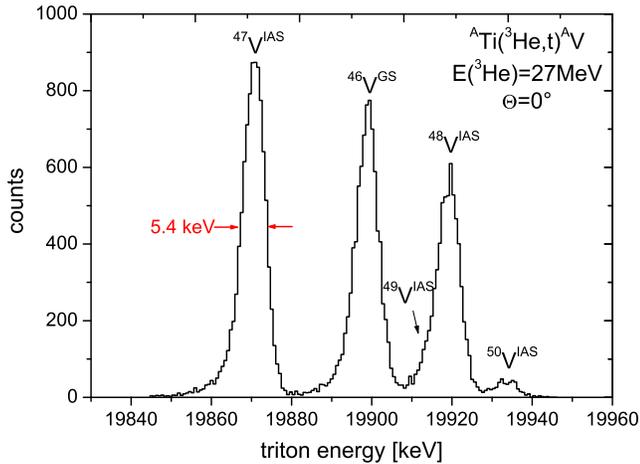}}
\caption
{Part of the ($^{3}$He,t) spectrum  for the mixed $^{46,47}$Ti target.} 
\end{figure}

An intense beam from the ECR like ion source \cite{Her} was accelerated in the Munich MP 
tandem accelerator to an energy of $27~MeV$. Typical beam currents of
$^3$He$^{2+}$ ions 
on target were $0.8~\mu A$. The beam energy was chosen as a compromise for 
optimum energy resolution. To minimize the specific energy loss a high energy 
would be favoured. On the other hand the relative energy resolution achieved 
with a magnetic spectrograph is constant and therefore calls for a low ion 
energy.
The triton energy from the $^{46}$Ti($^{3}$He,t)$^{46}$V reaction was 
(as in \cite{Von77}) measured with the Munich Q3D magnetic spectrograph, that 
provides  a superb intrinsic resolution of about $2\cdot 10^{-4}$. 
In the focal plane the tritons were identified and
their position was measured by a proportional counter with 
individual readout of 256 cathode strips \cite{Wir}. The
$^{46}$Ti($^{3}$He,t)$^{46}$V Q-value was calibrated against that of the 
$^{47}$Ti($^{3}$He,t)$^{47}$V reaction to the IAS in the T=3/2 multiplet of 
A=47. The difference of these Q-values is just equal to
the difference in Coulomb displacement energies (CDE) 
for the isotopes and thus small ($\approx 30~keV$). 
To become independent of effects of beam position 
on the target and on different beam energies and energy losses we measured 
both reactions simultaneously in one single target. The target was produced by 
evaporating 20 $\mu g/cm^2$ of a mixture of enriched $^{46}$Ti and $^{47}$Ti onto 
a 4 $\mu g/cm^2$ carbon backing. 
The isotopic composition of this mixed and other reference targets is given in table 1.

The Q3D was positioned at $0 ^o$:
for the L=0 transitions of interest the cross section has a maximum 
and the energy of the tritons is to first order independent of the angle. The 
angular acceptance was restricted to $\pm 2.3 ^o$. This causes a maximum energy 
shift of 2.5 $keV$ and thus a low energy tail. Since the magnetic rigidity 
of  $27~MeV$ $^3$He$^{2+}$ is only 58\% of 
that for $20~MeV$ $^3$H$^+$, the background from scattered beam particles in the focal 
plane detector was tolerable. 

To measure the particle position in the focal 
plane, we used a multiwire gas proportional counter \cite{Wir}. A precise 
position information is obtained from the charges influenced on 3 to 7 of 
255 cathode strips with a periodicity of $3.5~mm$. Every strip is provided with a 
preamplifier and shaper. The digitized charge 
signals are read out for every event and in the offline analysis fitted to a 
Gaussian to yield a position information with an intrinsic accuracy better 
than $0.1~mm$, corresponding to $10^{-5}$ in energy. For particle identification the 
energy loss signal of the proportional wire is used and a plastic scintillator 
yields the residual energy.
\begin{figure}[b,t]
\centerline{\includegraphics[width=9.5cm,angle=0]{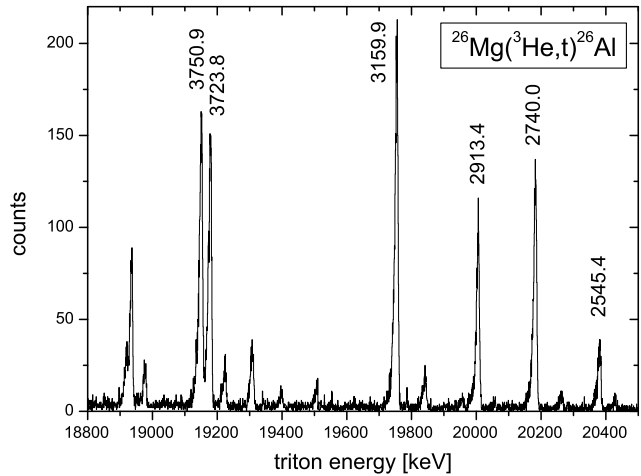}}
\caption
{Calibration spectrum $^{26}$Mg($^{3}$He,t)$^{26}$Al measured at the same beam energy and 
spectrograph setting. The states in $^{26}$Al used for the 
calibration are denoted with their excitation energy.} 
\end{figure}
\begin{figure}[b,t]
\centerline{\includegraphics[width=9.5cm,angle=0]{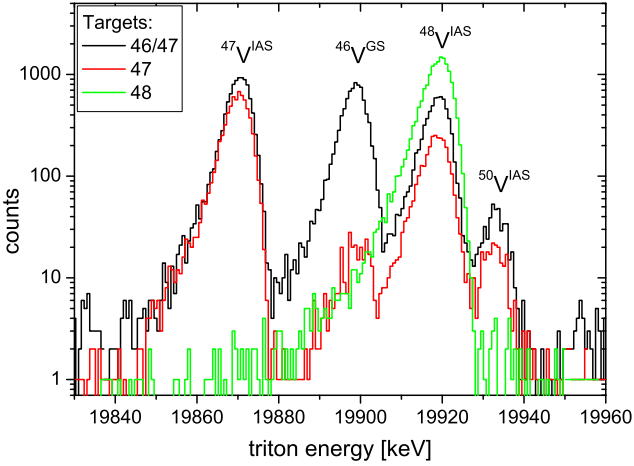}}
\caption
{Part of the ($^{3}$He,t) spectrum for enriched targets of $^{47}$Ti (red/grey)
$^{48}$Ti (green/light grey) and the mixed $^{46,47}$Ti target (black, same as Fig. 1).} 
\end{figure}

\section{Results and Discussion}
Fig. 1 shows the relevant part of the spectrum. The excitation of the  
IAS of all stable Ti isotopes is visible with intensities consistent with the 
respective isotopic content of the target and the expected 
neutron number dependence of the cross section $\sigma \propto (N-Z)/A^2$ \cite{Bec71}. 
The width of the peaks (FWHM) corresponds to $5.4~keV$, whereas the
resonance curves in the direct mass measurements \cite{Sav05,eron06}, 
inspite of the tremendous relative resolution,
have a width of about $30~keV$ and $50~keV$ (FWHM) respectively.
To determine the energy difference between the 
$^{46}$V and $^{47}$V peaks we need only the slope of the energy calibration
which we obtained from the $^{26}$Mg($^{3}$He,t)$^{26}$Al reaction under exactly the
same conditions. The spectrum is shown in Fig. 2.
With states between 2.5 and 3.8~$MeV$ of excitation
the quadratic relation between position and triton energy was fitted 
and the slope of the calibration was obtained with an uncertainty of 
less than $0.2\%$.

Thus we obtain a difference in triton energies for the $(^{3}$He,t) reactions 
on $^{46}$Ti and $^{47}$Ti of $28.27(16)~keV$. 
This results in a Q-value difference of $\Delta Q(46,47)=(28.73\pm 0.16)~keV$ due to
the different recoil energies of the heavy reaction products. 
The error consists of the uncertainty in the fitted peak positions 
and that of the energy calibration.
It has to be noted that in contrast to the measurement of Ref. \cite{Von77} 
systematic uncertainties like change in beam energy and position of beam or 
target
do not have to be considered, because of the simultaneous measurement.
For completeness we also give the triton energy difference of the pair $^{46}$Ti and $^{48}$Ti 
as $20.28(0.20)~keV$. The Q-value difference then is $\Delta Q(46,48)=(-18.57\pm 0.20)~keV$

One possible source of systematic errors could be due to unobserved lines
underneath the IAS. Therefore we also investigated the ($^{3}$He,t) reactions on
enriched $^{47}$Ti and $^{48}$Ti targets. The spectra are shown in Fig. 3 together with that 
of the mixed $^{46,47}$Ti target. The spectra are all consistent with the expectation from the 
known isotopic impurities. Even if there would be a $^{47}$V  or $^{48}$V line hidden under the $^{46}$V peak,
it could at most have a 0.8 \% contribution and could shift the $^{46}$V line
by at most $80~eV$. Thus we can neglect such an influence.
\begin{figure}[t]
\centerline{\includegraphics[width=9.5cm,angle=0]{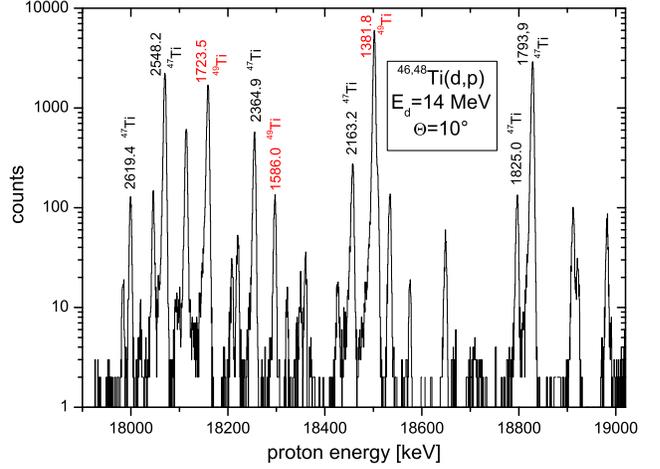}}
\caption
{Spectrum of the $^{46,48}$Ti(d,p)$^{47,49}$Ti reaction to determine
$S_n(^{47}$Ti). Excited states used for the adjustment in either 
$^{47}$Ti or $^{49}$Ti are denoted.} 
\end{figure}

We also considered differential nonlinearities in the position determination.
Since the particle position is derived from a number of 3 - 7  cathode strips,
such an effect is conceivable.
This was investigated by producing a white spectrum with scattered 
particles from a thick target.
Indeed a modulation of intensity was observed when determining the position of
a particle from the centroid of the pulse height distribution, which could
shift a line by at most $140~eV$. This effects can easily be understood
as caused by the threshold on individual strip signals. But our analyzed spectra were produced
by applying an event-by-event fitting of the peak position. With that procedure
no modulation of the white spectrum was observed and a peak shift cannot be more than $10~eV$.

The main uncertainty of the reference Q-value for the 
$^{47}$Ti$(^{3}$He,t)$^{47}$V reaction in the mass table \cite{Aud03} is
due to the neutron separation energy $S_n$ of $^{47}$Ti with  $S_n=8880.29$ $(0.29)~keV$. 
Therefore we measured the $^{46,48}$Ti(d,p)$^{47,49}$Ti reactions 
at a deuteron energy of $14~MeV$ with an enriched $^{46}$Ti target that
still contains $23\%$ of $^{48}$Ti. 
The spectrum is shown as Fig. 4.
Thus the Q-value difference between
the $^{46,48}$Ti(d,p)$^{47,49}$Ti reactions was determined 
with lines at excitation energies around $2.2~MeV$ in
$^{47}$Ti and $1.5~MeV$ in $^{49}$Ti to be $(S_n(^{47}$Ti$)-S_n(^{49}$Ti$))=(738.15\pm 0.25)~keV$
The uncertainty is mainly due to the uncertainty of the states in $^{47}$Ti $(\approx~0.2~keV)$
\cite{NDS47} and their scatter around our calibration of about $0.30~keV$ (rms).
With the averaged value $S_n(^{49}$Ti$)=(8142.358\pm0.013)~keV$ \cite{Aud03,Fir07}
we obtain $S_n(^{47}$Ti$)=(8880.51\pm0.25)~keV$. 
This is 
in good agreement with the literature value \cite{Aud03} that
already contains the new neutron capture value $8880.50(0.30)$ $keV$ \cite{Fir07}. 
Since all earlier data have much larger errors
we omit them and use the weighted average of the latter and our result: 
$S_n(^{47}$Ti$)=(8880.51\pm0.19)~keV$.

From Esch et al. \cite{Esc84} we have both the excitation energy of the IAS in $^{47}$V,
$E_x=(4150.35\pm0.11)~keV$, and the proton separation energy  
$S_p(^{47}$V$)=(5167.57\pm0.07)~keV$. 
The latter value needs a little adjustment, because it had been measured 
with a $^{46}$Ti(p,$\gamma )^{47}$V resonance
relative
to a resonance in the $^{27}$Al(p,$\gamma )^{28}$Si reaction.
That resonance energy has been remeasured since and Hardy et al. \cite{hardy06} use a new average 
for the proton energy $E_p=991.780(60)~keV$ instead of $E_p=991.880(40)~keV$
used in \cite{Esc84}.
Therefore we use the value $S_p(^{47}$V$)=(5167.67\pm0.09)~keV$.

Then we arrive at 
\begin{align*}
\begin{split}
Q_{EC}(\rm{^{46}V}) &=-\Delta Q(46,47)-S_p(\rm{^{47}V})+E_x(IAS,\rm{^{47}V})+\\
& \quad +S_n(\rm{^{47}Ti})+(m(\rm{H})-m(\rm{n}))\cdot c^2\\
& =(7052.11\pm0.27)~keV 
\end{split}
\end{align*}

This value is slightly smaller (2$\sigma$) and 
as precise as those from the direct mass measurements 
$(7052.90\pm0.40)$ $keV$ \cite{Sav05} and 
$(7052.72\pm0.31$ $keV$\cite{eron06}. The old 
reaction Q-value \cite{Von77} is still off by nearly 3$\sigma$. 
For the weighted average of all measurements we obtain
$Q_{EC}(^{46}$V$)=(7052.32\pm0.18)~keV$ and
$Q_{EC}(^{46}$V$)=(7052.49\pm0.18)~keV$ if we exclude the Vonach et al. value.
This shows that the value of Ref. \cite{Von77} has only little influence on the final 
result. 
\begin{figure}[t]
\centerline{\includegraphics[width=8.5cm,angle=0]{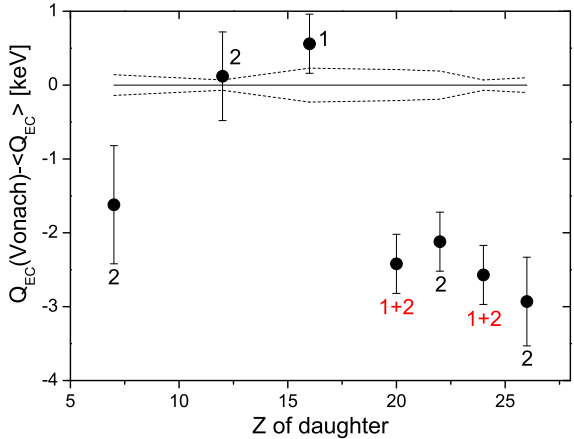}}
\caption
{
Comparison of the Q-values from \cite{Von77} with all other measurements.
The errors of the latter are indicated by the dashed line at $\Delta Q=0$.
The numbers 1 and 2 indicate the two measuring series of \cite{Von77}, "1"
was measured relative to the TOF system. In series "2" the A=42 and A=50 
Q-values served as reference.
} 
\end{figure}

In Fig. 5 we show a comparison between all Q-values from \cite{Von77} 
with all other measurements (\cite{{eron08},{hardy05},{Sav05},{eron06}}
and this work). The four values for the A=42, 46, 50, 54 systems 
deviate by $5 \sigma$. There are also large deviations within 
either measuring series pointing to systematic deviations
not taken into account in the uncertainties of \cite{Von77}. 
Therefore we think it is justified to
completely ignore all data of Vonach et al \cite{Von77}.
The new average $Q_{EC}(^{46}$V$)=(7052.49\pm0.18)~keV$ is
$0.31~keV$ smaller than without our measurement and lowers the 
$ft$-value by only $0.7~s$ compared to Ref. \cite{hardy08}, well within the error bars.

We have determined the electron capture decay Q-value for $^{46}$V as precise
and in reasonable agreement with recent mass measurements. 
This shows 
that reaction Q-values are competitive
with direct mass measurements as long as systematic uncertainties are avoided. 
Although Penning trap mass measurements still have
potential to improve precision, a cross check with a completely 
independent method can increase the confidence.
We have used a novel method by calibrating simultaneously with
the same reaction on another target isotope and
exciting the IAS.
Thus most of the systematic errors are avoided.
The method in principle can be applied to other Q-values
of superallowed $\beta$ emitters.
However at the moment there is no other case for which both
the ground state Q-value and the energy of the IAS of another isotope is known
with sufficient precision.

\section{Acknowledgement}
Valuable discussions with J.C. Hardy are gratefully acknowledged.
We also thank the crew of the MLL tandem accelerator.
This work was supported by the DFG cluster of excellence
"Origin and Structure of the Universe" (www.universe-cluster.de).


\end{document}